%% file: MRRspiresed2.tex
\documentstyle[epsfig]{mn2e}
\begin{document}
\input BoxedEPS.tex
\input macro.tex
\SetEPSFDirectory{/scratch/sbgs/figures/hst/}
\SetRokickiEPSFSpecial
\HideDisplacementBoxes

\title[Cold dust and young starbursts]{Cold dust and young starbursts: spectral energy distributions 
of {\it Herschel} SPIRE sources from the HerMES survey
\thanks{{\it Herschel} is an ESA space observatory with science
instruments provided by European-led Principal Investigator
consortia and with important participation from NASA.}
}
\author[Rowan-Robinson M. et al]{M.~Rowan-Robinson$^{1}$\thanks{E-mail: \texttt{m.rrobinson@imperial.ac.uk}},
 I.G.~Roseboom$^{2}$,
M.~Vaccari$^{3}$,
\newauthor
 A.~Amblard$^{4}$,
 V.~Arumugam$^{5}$,
 R.~Auld$^{6}$,
 H.~Aussel$^{7}$,
T.~Babbedge$^{1}$,
\newauthor
 A.~Blain$^{8}$,
 J.~Bock$^{8,9}$,
 A.~Boselli$^{10}$,
 D.~Brisbin$^{11}$,
 V.~Buat$^{10}$,
 D.~Burgarella$^{10}$,
\newauthor
N.~Castro-Rodriguez$^{12}$,
 A.~Cava$^{12}$,
 P.~Chanial$^{1}$,
 D.L.~Clements$^{1}$,
 A.~Conley$^{13}$,
\newauthor
 L.~Convers$i^{14}$,
A.~Cooray$^{4,8}$,
 C.D.~Dowell$^{8,9}$,
 E.~Dwek$^{15}$,
S.~Dye$^{6}$,
S.~Eales$^{6}$,
\newauthor
 D.~Elbaz$^{7}$,
 D.~Farrah$^{2}$,
 M.~Fox$^{1}$,
 A.~Franceschini$^{3}$,
W.~Gear$^{6}$,
J.~Glenn$^{13}$,
\newauthor
 E.A.~Gonz\'alez~Solares$^{16}$,
 M.~Griffin$^{6}$,
 M.~Halpern$^{17}$,
\newauthor
 E.~Hatziminaoglou$^{19}$,
 J.~Huang$^{20}$,
 E.~Ibar$^{21}$,
 K.~Isaak$^{6}$,
R.J.~Ivison$^{21,5}$,
\newauthor
G.~Lagache$^{22}$,
L.~Levenson$^{8,9}$,
N.~Lu$^{8,23}$,
S.~Madden$^{7}$,
 B.~Maffei$^{24}$,
\newauthor
 G.~Mainetti$^{3}$,
 L.~Marchetti$^{3}$,
A.M.J.~Mortier$^{1}$,
H.T.~Nguyen$^{8,9}$,
\newauthor
 B.~O'Halloran$^{1}$,
 S.J.~Oliver$^{2}$,
A.~Omont$^{25}$,
 M.J.~Page$^{26}$,
 P.~Panuzzo$^{7}$,
\newauthor
 A.~Papageorgiou$^{6}$,
 H.~Patel$^{1}$,
 C.P.~Pearson$^{27,28}$,
 I.~Perez Fournon$^{12}$,
M.~Pohlen$^{6}$,
\newauthor
 J.I.~Rawlings$^{26}$,
 G.~Raymond$^{6}$,
 D.~Rigopoulou$^{27,29}$,
 D.~Rizzo$^{1}$,
B.~Schulz$^{8,23}$,
\newauthor
 Douglas~Scott$^{17}$,
N.~Seymour$^{26}$,
D.L.~Shupe$^{8,23}$,
A.J.~Smith$^{2}$,
 J.A.~Stevens$^{30}$,
\newauthor
M.~Symeonidis$^{26}$,
M.~Trichas$^{1}$,
K.E.~Tugwell$^{26}$,
I.~Valtchanov$^{14}$,
\newauthor
L.~Vigroux$^{25}$,
L.~Wang$^{2}$,
R.~Ward$^{2}$,
G.~Wright$^{21}$,
C.K.~Xu$^{8,23}$,
M.~Zemcov$^{8,9}$\\
${1}$Astrophysics Group, Imperial College London, Blackett Laboratory, Prince Consort Road, London SW7 2AZ, UK\\
% \email{mrr@imperial.ac.uk},
$^{2}$Astronomy Centre, Dept. of Physics \& Astronomy, University of Sussex, Brighton BN1 9QH, UK,\\
$^{3}$Dipartimento di Astronomia, Universit\`{a} di Padova, vicolo Osservatorio, 3, 35122 Padova, Italy,
$^{4}$Dept. of Physics \& Astronomy,\\ University of California, Irvine, CA 92697, USA,
$^{5}$Institute for Astronomy, University of Edinburgh, Royal Observatory, Blackford Hill,\\ Edinburgh EH9 3HJ, UK,
$^{6}$Cardiff School of Physics and Astronomy, Cardiff University, Queens Buildings, Cardiff CF24 3AA, UK,\\
$^{7}$Laboratoire AIM-Paris-Saclay, CEA/DSM/Irfu - CNRS - Universit\'e Paris Diderot, CE-Saclay, pt courrier 131, F-91191\\ Gif-sur-Yvette, France,
$^{8}$California Institute of Technology, 1200 E. California Blvd., Pasadena, CA 91125, USA,\\
$^{9}$Jet Propulsion Laboratory, 4800 Oak Grove Drive, Pasadena, CA 91109, USA,
$^{10}$Laboratoire d'Astrophysique de Marseille,\\ OAMP, Universit\'e Aix-marseille, CNRS, 38 rue Fr\'ed\'eric Joliot-Curie, 13388 Marseille cedex 13, France,
$^{11}$Space Science Building,\\ Cornell University, Ithaca, NY, 14853-6801, USA,
$^{12}$Institute de Astrofisica de Canarias,  C/ Via Lactea s/n, E-38200\\ La Laguna, Tenerife,
Spain,
$^{13}$Departamento de Astrof{\'\i}sica, Universidad de La Laguna (ULL), E-38205 La Laguna, Tenerife, Spain\\
$^{14}$Dept. of Astrophysical and Planetary Sciences, CASA 389-UCB, University of  Colorado, Boulder, CO 80309, USA,\\
$^{15}$Herschel Science Centre, European Space Astronomy Centre, Villanueva de la Ca\~nada,  28691 Madrid, Spain,\\
$^{16}$Observational  Cosmology Lab, Code 665, NASA Goddard Space Flight  Center, Greenbelt, MD 20771, USA,
$^{17}$Institute of Astronomy,\\ University of Cambridge, Madingley Road, Cambridge CB3 0HA, UK,
$^{18}$Department of Physics \& Astronomy, University of British\\ Columbia, 6224 Agricultural Road, Vancouver, BC V6T~1Z1, Canada,
$^{19}$ESO, Karl-Schwarzschild-Str. 2, 85748 Garching bei M\"unchen, Germany,
$^{20}$Harvard-Smithsonian Center for Astrophysics, \\ MS65, 60 Garden Street,  Cambridge,  MA02138, USA,
$^{21}$UK Astronomy Technology Centre, Royal Observatory, Blackford Hill, \\ Edinburgh EH9 3HJ, UK,
$^{22}$Institut d'Astrophysique Spatiale (IAS), b\^atiment 121, Universit\'e Paris-Sud 11 and CNRS (UMR 8617),\\  91405 Orsay, France,
$^{23}$Infrared Processing and Analysis Center, MS 100-22, California Institute of Technology, JPL, Pasadena,\\ CA 91125, USA,
$^{24}$School of Physics and Astronomy, The University of Manchester, Alan Turing Building, Oxford Road,\\ Manchester M13 9PL, UK,
$^{25}$Institut d'Astrophysique de Paris, UMR 7095, CNRS, UPMC Univ. Paris 06, 98bis boulevard Arago,\\ F-75014 Paris, France,
$^{26}$Mullard Space Science Laboratory, University College London, Holmbury St. Mary, Dorking,\\ Surrey RH5 6NT, UK,
$^{27}$Space Science \& Technology Department, Rutherford Appleton Laboratory, Chilton, Didcot, Oxfordshire\\ OX11 0QX, UK,
$^{28}$Institute for Space Imaging Science, University of Lethbridge, Lethbridge, Alberta T1K 3M4, Canada,\\
$^{29}$Astrophysics, Oxford University,Keble Road, Oxford OX1 3RH, UK,
$^{30}$Centre for Astrophysics Research,\\ University of Hertfordshire, College Lane, Hatfield, Hertfordshire AL10 9AB, UK\\
}
\maketitle
\begin{abstract}
We present spectral energy distributions (SEDs) for 68  {\it Herschel} sources  
detected at 5-$\sigma$ at 250, 350
and 500 $\mu$m in the HerMES SWIRE-Lockman field.  
We explore whether existing models for starbursts, quiescent star-forming galaxies and for AGN dust tori 
are able to model the full range of SEDs measured with {\it Herschel}.  We find that while many galaxies 
($\sim 56 \%$) are well fitted with the templates used to fit {\it IRAS}, {\it ISO} and {\it Spitzer} sources, 
for about half the galaxies two new templates are required: quiescent ('cirrus')
models with colder (10-20 K) dust, and a young starburst model with higher optical depth than Arp 220.
Predictions of submillimetre fluxes based on model fits to 4.5-24 $\mu$m data agree rather poorly with 
the observed fluxes, but the agreement is better for fits to 4.5-70 $\mu$m data.  {\it Herschel} galaxies detected at 500 $\mu$m tend to
be those with the very highest dust masses.

\end{abstract}

%\large

\section{Introduction}
The combination of {\it Herschel} (Pilbratt et al 2010) and {\it Spitzer} data provides us with the first 3-500 $\mu$m spectral energy distributions of large samples of galaxies, so that we can determine accurately the masses of cold dust present in a substantial sample of galaxies and search for very
young, heavily obscured starbursts.  The HerMES wide-area surveys (Oliver et al 2010a) have been targeted on fields in which we have excellent {\it Spitzer} data.

Over the past twenty years increasingly sophisticated radiative transfer models for different types of infrared galaxy
have been developed, for example for  starburst galaxies
(Rowan-Robinson \& Crawford 1989, Rowan-Robinson \& Efstathiou 1993, Silva et al 1998, Efstathiou et al 2000,
Takagi et al 2003, Siebenmorgen \& Krugel 2007),
AGN dust tori (Rowan-Robinson \& Crawford 1989, Pier \& Krolik 1992, Granato \& Danese 1994, Efstathiou \& 
Rowan-Robinson 1995, Rowan-Robinson 1995, Nenkova et al 2002, 2008, Fritz et al 2006, H\"{o}nig et al 2006, Schartmann et al 2008), 
and quiescent ('cirrus') galaxies (Rowan-Robinson 1992, Silva et al 1998, Dale et al 2001, Efstathiou \& Rowan-Robinson 2003, 
Dullemond \& van Bemmel 2005, Piovan et al 2006, Draine  \& Li 2006, Efstathiou 
\& Siebenmorgen 2009).  Each of these model types involves at least two significant model parameters 
so there are a great wealth of possible models, particularly as a galaxy SED may be a mixture of all three types.

Rowan-Robinson and Efstathiou (2009) have shown how these models can be used to understand the
 interesting diagnostic diagram of Spoon et al (2007) for starburst and active galaxies, which
plots the strength of the 9.7 $\mu$m silicate feature against the equivalent width of the 6.2 $\mu$m PAH feature
for 180 galaxies with Spitzer IRS spectra.  Increasing depth of the 9.7 $\mu$m silicate feature is, broadly, a measure of the youth of the starburst,
because initially the starburst is deeply embedded in its parent molecular cloud.  The detailed starburst model of
Efstathiou et al (2000) shows the evolution of the starburst SED through the whole history of the starburst, from the
deeply embedded initial phase through to the Sedov expansion phase of the resulting supernovae.
However Rowan-Robinson and Efstathiou did find that there was some aliasing between young starbursts and heavily obscured AGN: the submillimetre data of {\it Herschel} can help to break this ambiguity, since young starbursts are
expected to be much more prominent in the far infrared and submillimetre than AGN dust tori.

Often, however, we have only limited broad-band data available and in this situation it is more illuminating
to use a small number of infrared templates to match the observed infrared colours (eg Rowan-Robinson
and Crawford 1989, Rowan-Robinson 1992, 2001, Rowan-Robinson and Efstathiou 1993,
Rowan-Robinson et al 2004, 2005, 2008).   A set of just four templates (a quiescent 'cirrus' component, M82- and
Arp 220-like starbursts, and an AGN dust torus model) 
have proved remarkably successful in matching observed {\it ISO} and Spitzer SEDs, including cases where 
Spitzer Infrared Spectrograph (IRS) data are 
available (Rowan-Robinson et al 2006, Farrah et al 2008, Hernan-Caballero et al 2009).
In this paper we explore whether this simple four-template approach works for galaxies detected
by the SPIRE array (Griffin et al 2010) on {\it Herschel}, and what additional infrared components may be present.  

A cosmological model with $\Lambda$=0.7, $h_0$=0.72 has been used throughout.

\section{Selection of sample with good quality fluxes at 24, 250, 350 and 500 $\mu$m}

In this analysis we have focused on early HerMES\footnote{hermes.sussex.ac.uk}  (Oliver et al 2010a) Science Demonstration 
data in the SWIRE-Lockman area, where we  have optical 
and 3.6-160 $\mu$m photometry, photometric redshifts, and infrared template fits from the SWIRE photometric redshift catalogue 
(Rowan-Robinson et al 2008).  Obviously we are particularly interested in whether any new galaxy populations can be 
seen in the {\it Herschel} data.  

Photometry in the SPIRE bands was carried out via a linear inversion of the 
SPIRE maps, using the positions of known 24 $\mu$m sources and the SPIRE point-source response function (PSF) as 
an input. The PSF was assumed to be a Gaussian with full width to half power of 
18.2, 25.2 and 36.3 arcsec at 250, 350 and 500 $\mu$m respectively. 
The 24 $\mu$m input catalogue was optimized to 
alleviate concerns about overfitting. The method and description of the 
catalogue are presented in Roseboom et al. (2010).
In the 5 sq deg SWIRE-
Lockman area, 5225 sources were detected at 5-$\sigma$ in at least one band and 2367 of these are associated with sources in the 
SWIRE photometric redshift catalogue.  We have focused on the 70 sources detected at
5-$\sigma$ in all three SPIRE bands, which also satisfy some further restrictions.  Specifically, we use the recommended selections
described in Roseboom et al. (2010) which ensure robust solutions from the inversion process, i.e. low $\chi^2$ and
minimal correlations with neighbouring sources. This is essentially a 500 $\mu$m selected sample, with a flux-limit $\sim$27mJy.  
The combination of PSF fitting, the choice of 24 $\mu$m targets only, and the elimination of confused
sources, allows us to reach fainter fluxes than would be possible in an unbiassed survey.
 Our requirement of association with a SWIRE 24 $\mu$m source discriminates against sources with
S(500)/S(24) $>$ 200, and our requirement of an entry in the SWIRE Photometric Redshift Catalogue discriminates
against sources with {\it z}$>$ 1.5.  Selection at 500 $\mu$m favours galaxies with cooler dust, than say selection at
70 or even 250 $\mu$m.
From HerMES counts at 500 $\mu$m (Oliver et al 2010b) we deduce that there should be 124$\pm$16 500 $\mu$m sources
brighter than this flux in the 5 sq deg area of this study.  Thus our sample represents 46-66$\%$ of the total 500 $\mu$m
population.  The remaining sources are presumably fainter than the the limit of the SWIRE optical photometry (r$\sim$25)
or the SWIRE 24 $\mu$m limit (S(24) $\sim$ 100$ \mu$Jy).
Since confusion is an issue, particularly at 500 $\mu$m, we have carefully examined all
24-$\mu$m sources within 40" of SPIRE sources to assess whether these neighbours could have contributed significantly to the 500 $\mu$m
flux.  We eliminated 2 of the 70 sources as having neighbours likely to have contributed $> 30 \%$ of the 500 $\mu$m flux.  

\section{Spectral energy distributions of {\it Herschel} galaxies}
We have modelled the SEDs of the remaining 68 3-band sources, following the methodology of Rowan-Robinson et al (2008), 
and the results are presented in Figs 1-6, in redshift order.  Optical and near infrared data are fitted with one of six galaxy
templates and two QSO templates, with the extinction $A_V$ as a free parameter.  Infrared and submillimetre data are
fitted initially with a combination of four infrared templates (cirrus, M82 and A220 starbursts, or AGN dust torus).
Parameters of the fits are given in Table 2.  Some unpublished spectroscopic redshifts were supplied by Huang, Rigopoulou et al
(in preparation).  
Photometric redshifts have been indicated in Figs 1-6 and Table 2 with brackets.  The accuracy of these 
redshifts is $\sim5\%$ in (1+z) for most of the galaxies, where 4 or more photometric bands are available
(Rowan-Robinson et al 2008).  The full $\chi^2$ distribution for the photometric redshifts is given in
the SWIRE Photometric Redshift Catalogue.
The optical galaxy templates are those of Rowan-Robinson et al (2008) and are shown at full resolution
in the SED plots.  The optical types given in Table 2 are optical template types only and contain no
morphological information.  None of the 68 objects are radio-loud AGN and none show evidence for
non-thermal emission at 3.6-500 $\mu$m.

\begin{figure*}
\epsfig{file=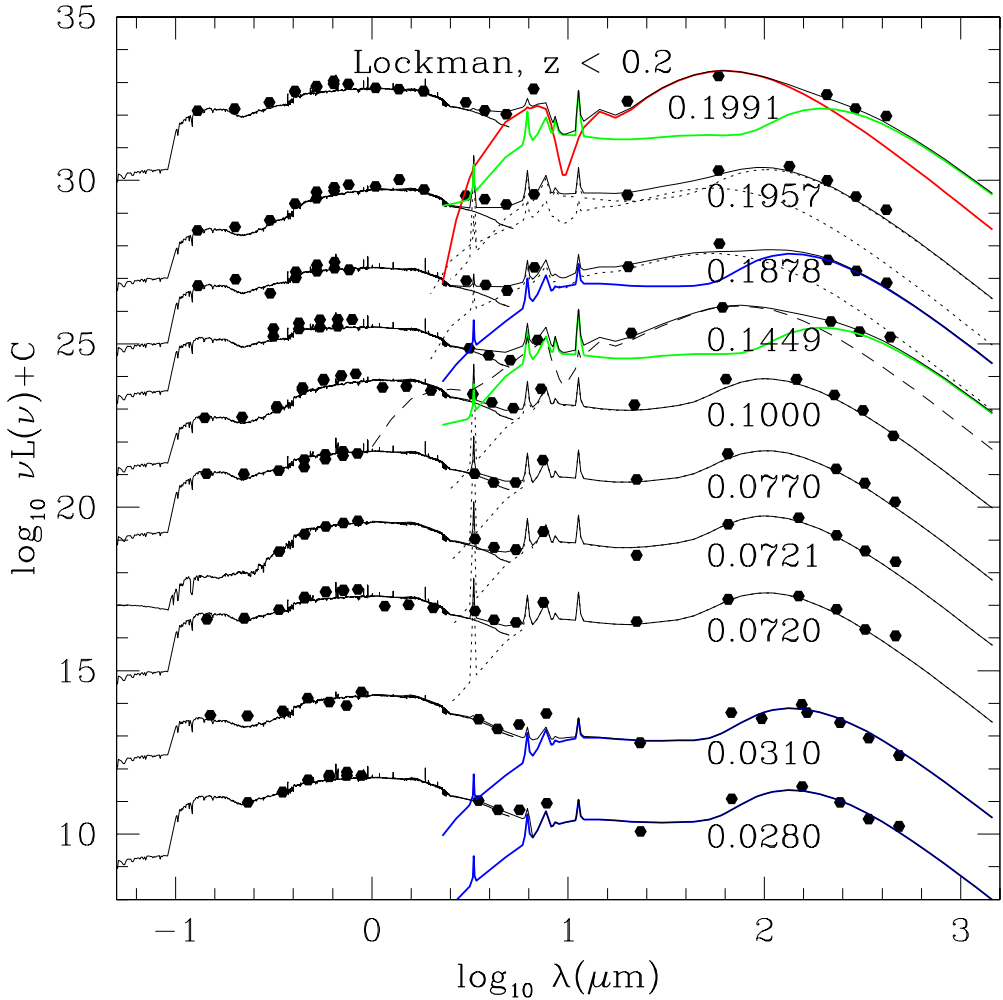,angle=0,width=14cm}
\caption{SEDs for SWIRE-Lockman 3-band galaxies with {\it z}$<$ 0.2, labelled with their redshift (all are spectroscopic).   Blue and green 
curves are quiescent ('cirrus') model with colder dust ($\psi$ = 1, 0.1).  The red curve is a young starburst model. 
}
\end{figure*}

While the four standard infrared templates work well for many sources, the 350 and 500 
$\mu$m fluxes often require the presence of colder dust than is incorporated into our four basic templates.  
The two new templates used here are taken from the range of optically thin interstellar medium ('cirrus') templates developed by 
Rowan-Robinson (1992) and Efstathiou and Rowan-Robinson (2003).  The key parameter determining the temperature of the 
dust grains is the intensity of the radiation field, which we can characterize by the ratio of intensity of 
radiation field to the local interstellar radiation field, $\psi$.  The standard cirrus template corresponds to $\psi$ = 5, and this
is the value used by Rowan-Robinson (1992) to fit the central regions of our Galaxy.  $\psi$ = 1 corresponds to the interstellar
radiation field in the vicinity of the Sun.  We also find that some galaxies need a much lower intensity radiation field than this,
with $\psi$ = 0.1.  The corresponding grain temperatures in the dust model of Rowan-Robinson (1992) are given in Table 1.  For the
two new templates, the range of dust grain temperatures are 14.5-19.7 K and 9.8-13.4 K respectively.  Full details of the
 templates used are given at http:$//$astro.ic.ac.uk$/\sim$mrr$/$spire$/$templates

The need for cooler dust templates can also be seen clearly in a plot of S(500)/S(24) versus
redshift (Fig 7), in which the predictions of different templates are shown.  At  {\it z}$<$ 1, a significant 
fraction of galaxies require colder dust than the standard cirrus model.  Hints of this population were seen at {\it z}$<$ 0.4 in
the plot of {\it ISO} 175$/$90 $\mu$m flux ratio versus redshift (Fig 23) of Rowan-Robinson et al (2004).  Symeonidis et al (2009)
plotted a very similar figure, 160$/$70 $\mu$m flux ratio versus redshift, for Spitzer data.  They interpreted this as implying 
strong evolution in the cold dust component.

Eighteen of the 68 sources modelled in Figs 1-6 are '350 $\mu$m peakers' (S(250) $<$ S(350), S(500) $<$ S(350)). 
All are at redshift $>$ 0.9 and 10 have {\it z}$>$ 1.5.
Six, which are given as lower redshift ($<$0.5) in the SWIRE photometric redshift catalogue, clearly require higher redshift 
to fit their SEDs.   In each of these cases the photometric redshift was based on only two photometric bands, so of very low
reliability.  The adopted redshifts for these six galaxies are indicated in brackets in Table 1 with only two significant figures.  
The remaining photometric redshifts appeared plausible from the SED fits.
5 of the 68 sources are '500 $\mu$m peakers' (S(250) $<$ S(350) $<$ S(500)): all of these are at {\it z}$>0.9$ and 3 are 
at {\it z}$>$ 1.5.  So 350 and 500$\mu$m peakers are a reasonably good indication of high redshift.  However the range of infrared 
template required makes any determination of redshift from {\it Herschel} data alone problematic.  The six templates  $\psi$=0.1 cirrus, 
$\psi$=1 cirrus, $\psi$=5 cirrus,
Arp 220 starburst, t=0 starburst, and M82 starburst, have their  $\nu S_{\nu}$ peaks at 202, 136, 102, 73, 63 and 53$\mu$m, 
respectively, which could give rise to a range of a factor 3.8 in the determined (1+z). 

For a few sources plotted in Figs 1-6, the 500 $\mu$m point lies higher than the template fits.  
This is probably due to residual effects of confusion at 500 $\mu$m. Four sources in Figs 6 and 8
have observed fluxes highly discrepant with the models, three of which are {\it Spitzer} 70 $\mu$m fluxes.
These require further investigation.

Since only one of the 3-band sample has {\it z}$>$3, we have selected all SWIRE-Lockman galaxies with z$>$3 
detected at 5-$\sigma$ at 250 and 350 $\mu$m.  There are 15 of them (including the 3-band source) and Fig 8 shows 
SEDs for 11 of them.  500 $\mu$m fluxes are included ifthey are better than 3-$\sigma$.  9 are 
hyperluminous ($L_{ir}>10^{13} L_{\odot}$) M82-like starbursts, 4 of these with AGN dust tori, and just 2 are Arp220-like 
starbursts.  None of these {\it z}$>$ 3  galaxies show evidence of the cold dust components seen at lower redshift but this 
is an effect of the redshift, since the rest-frame wavelength corresponding to the SPIRE bands is at $\le$100 $\mu$m.  
We show a young starburst fit for one of the objects (see below).  It would be an acceptable alternative to the M82 
starburst fit for this object and perhaps for a couple of others.  Additional photometry will clarify this ambiguity.  The
other templates are reasonably distinct in their peak wavelength (see Table 1) so less susceptible to aliasing.
Where the 500 $\mu$m fluxes are less than 3-$\sigma$ we have checked that the model fits are consistent with
the 3-$\sigma$ limits.

\begin{table*}
\caption{Colour temperature, dust grain temperatures, and peak wavelengths of the templates used here}
\begin{tabular}{lllll}
template & '$T_d$' & actual dust & peak $\lambda$ & 60 $\mu$m bol.\\
 & ($\nu^2 B_{\nu}(T_d)$ & temperature & & correction\\
& (K) & (K) &  ($\mu$m)  &\\
& & & & \\
cirrus $\psi$=0.1 & 12 & 9.8-13.4 & 202 &10.9 \\
cirrus $\psi$=1 & 17.5 & 14.5-19.7 & 136 &8.38\\
cirrus $\psi$=5 & 23.5 & 19.1-24.1 & 102 &3.30\\
Arp220 sb & 33 & 3-1000 & 73 & 1.50\\
young sb (t=0) & 38 & 3-1000 & 63 & 1.28\\
M82 sb & 45 & 3-1000 & 53 & 1.75\\
\end{tabular}
\end{table*}

To search for young starbursts, previously indicated by IRS spectra with very deep silicate features
(Rowan-Robinson and Efstathiou 2009), we show a plot of S(70)/S(24) versus redshift for the 360 sources detected
at 5-$\sigma$ at 70, 250 and 350 $\mu$m, compared with the predictions of existing templates (Fig 9).  There is a clear population of
sources showing deeper 10 $\mu$m absorption than the Arp 220 template and we have modelled the SEDs of
a selection of these in Fig 10.  All 9 sources have very similar SEDs, with warm 100-500 $mu$m colours but relatively weak 24 $\mu$m fluxes,
and are well-fitted with a very young starburst model.  Sources in the redshift range
1.35-1.45 for which 24 $\mu$m would fall in the deep silicate absorption, would be discriminated against by their
faintness at 24 $\mu$m.  To confirm the reality of this component we need to obtain SPIRE photometry of sources
identified as having very deep silicate features from Spitzer IRS observations.  To identify these young starbursts we
need both {\it Spitzer} and {\it Herschel} data.  They can be masked by additional cirrus emission in the galaxy, as in the case of
161.34665+57.51625 in Fig 1.

Figure 11 shows infrared luminosity versus redshift for all
the infrared components listed in Table 2.  If a galaxy is fitted with several components, each is shown separately here.
Cirrus components, including the new colder templates, are seen at redshifts out to
1.7 and at luminosities up to $\sim 10^{12} L_{\odot}$.  Higher redshift galaxies may also have cool dust components, 
but we would need photometry at $\lambda > 500 \mu m$ to characterize them.
Higher luminosity sources are generally M82 or A220 starbursts, but the latter
do not dominate amongst hyperluminous ($L_{ir} > 10^{13} L_{\odot}$ galaxies.  The young starbursts are seen at {\it z}= 0.2-1.1
and are amongst the highest luminosity objects in their redshift range, but this is partially a selection
effect because of the requirement of a 70 $\mu$m detection.   Young starbursts may also be present in higher
redshift objects (see eg Fig 8), where additional photometry would be needed to eliminate aliasing with an M82 starburst. 

\begin{figure}
\epsfig{file=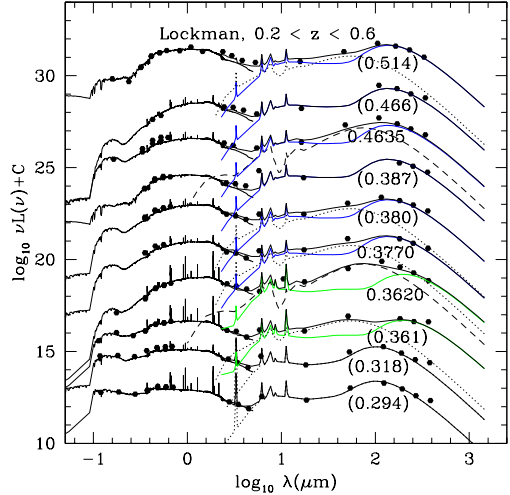,angle=0,width=7cm}
\caption{SEDs for SWIRE-Lockman 3-band galaxies with 0.2 $<$ {\it z}$<$ 0.5.  Photometric redshifts are indicated with brackets.
}
\end{figure}

\begin{figure}
\epsfig{file=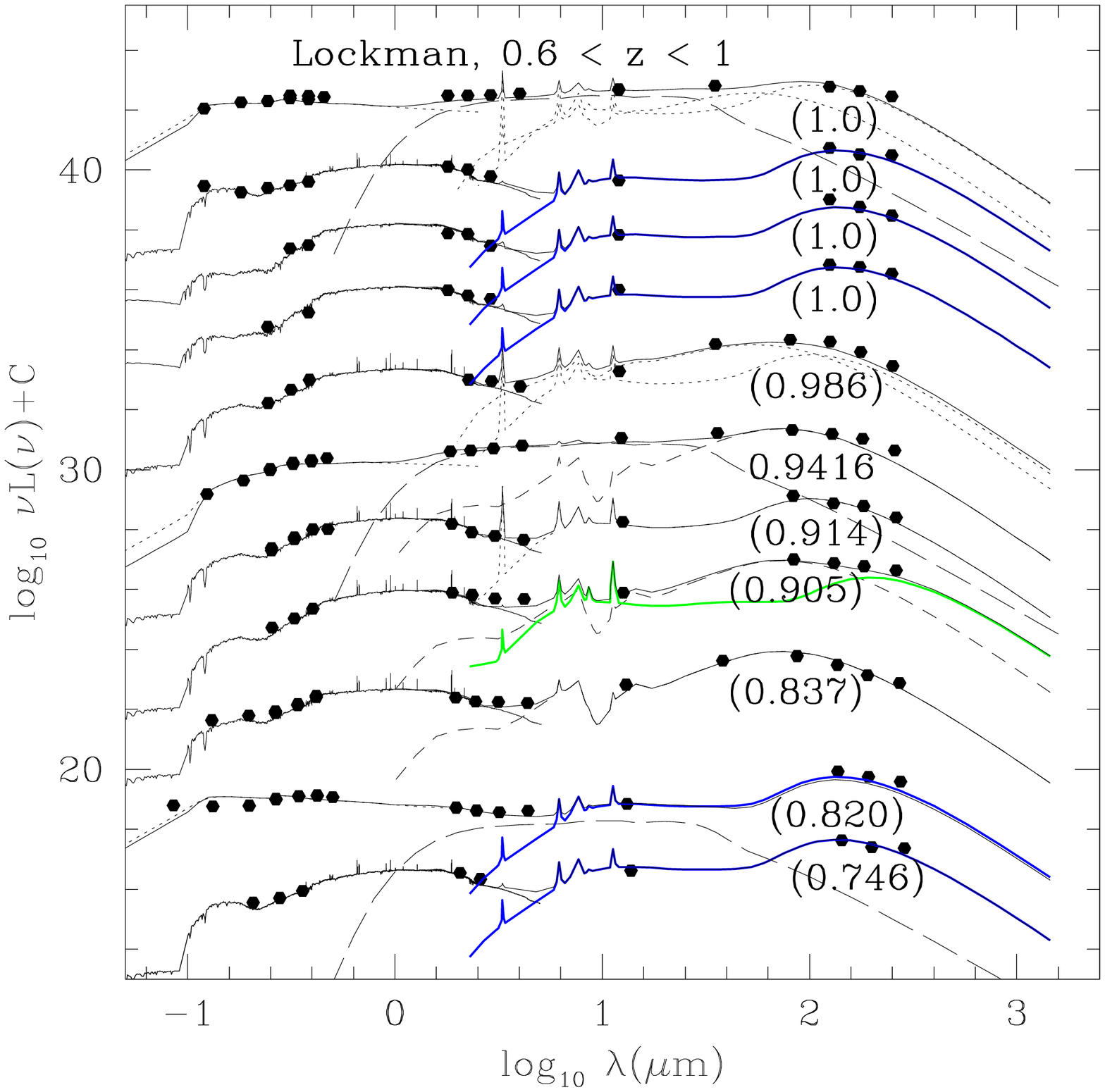,angle=0,width=7cm}
\caption{SEDs for SWIRE-Lockman 3-band galaxies with 0.5 $<$ {\it z}$<$ 1. 
}
\end{figure}

\begin{figure}
\epsfig{file=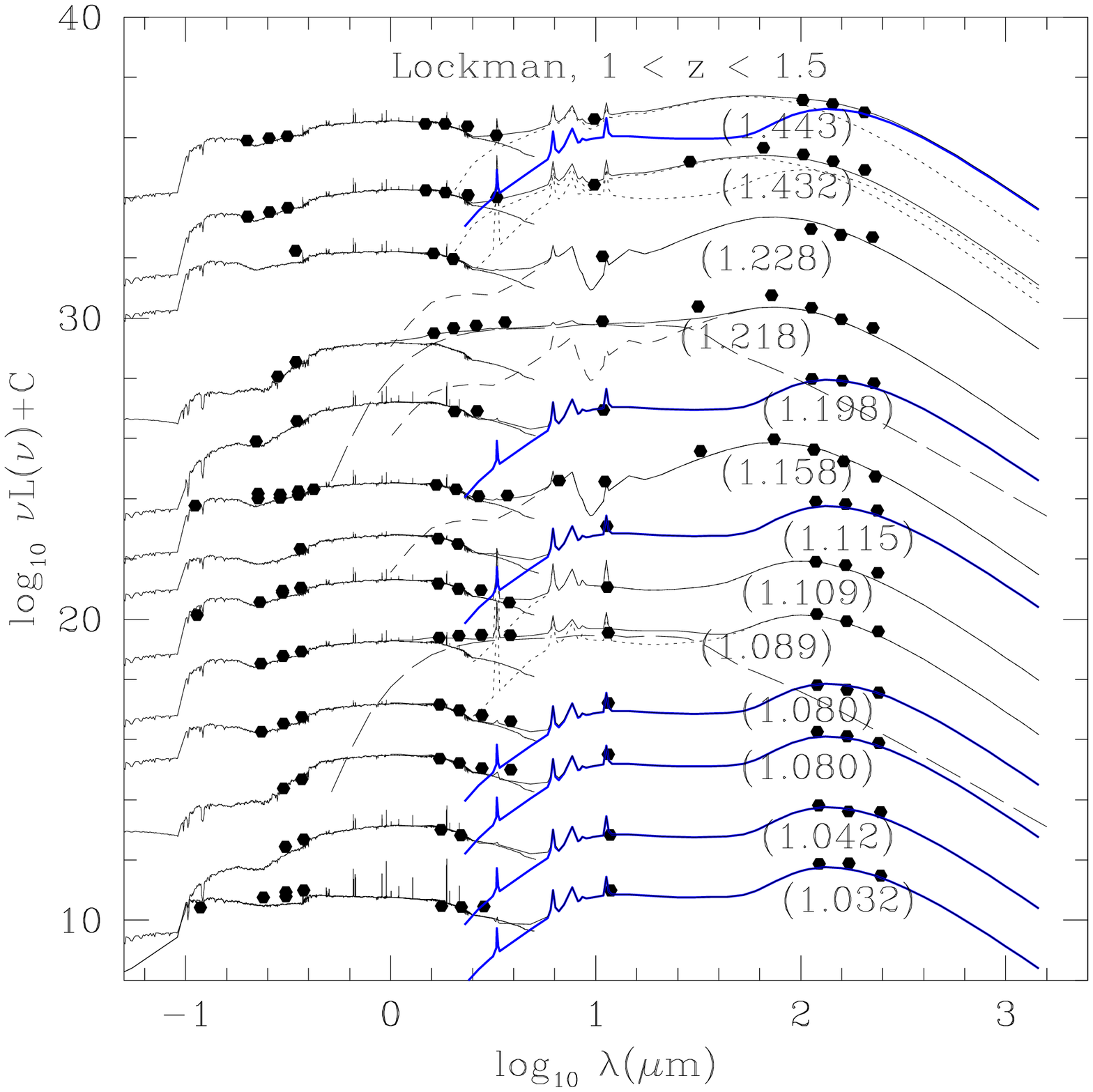,angle=0,width=7cm}
\caption{SEDs for SWIRE-Lockman 3-band galaxies with 1 $<$ {\it z}$<$ 1.5. 
}
\end{figure}

\begin{figure}
\epsfig{file=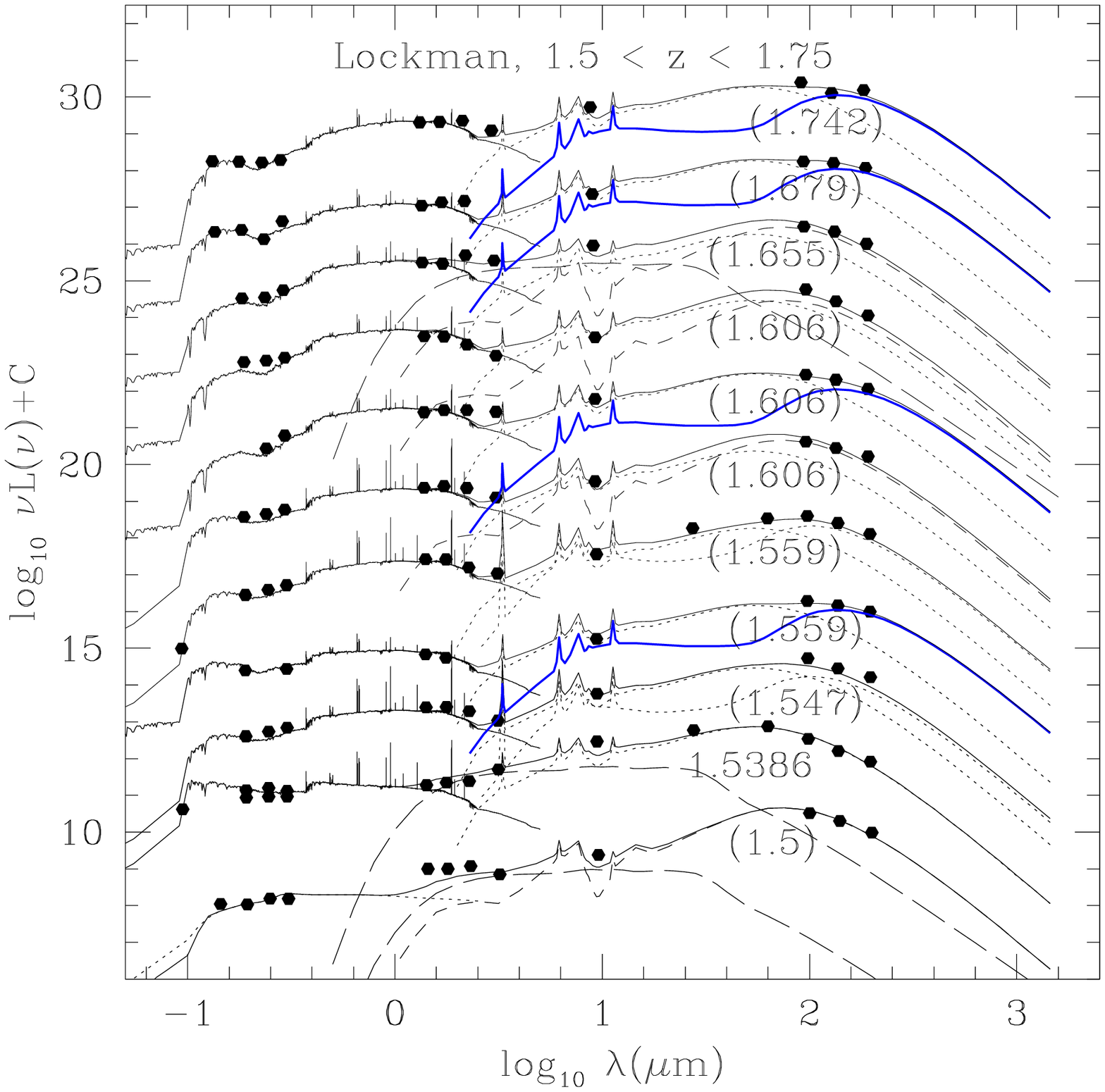,angle=0,width=7cm}
\caption{SEDs for SWIRE-Lockman 3-band galaxies with 1.5 $<$ {\it z}$<$ 1.75. 
}
\end{figure}

\begin{figure}
\epsfig{file=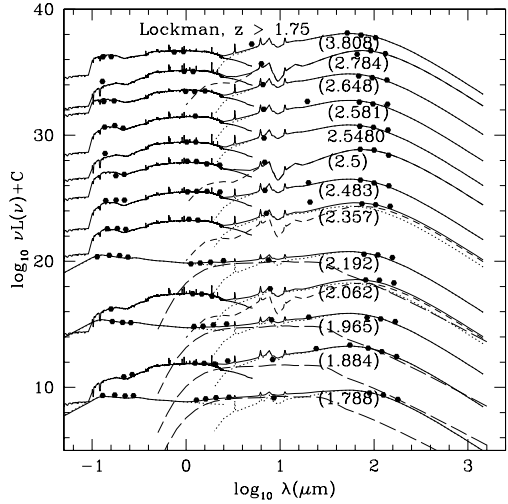,angle=0,width=7cm}
\caption{SEDs for SWIRE-Lockman 3-band galaxies with {\it z}$>$ 1.75. 
}
\end{figure}

We can compare the predicted submillimetre fluxes from the SWIRE Photometric Redshift Catalogue template fits
with the observed {\it Herschel} fluxes.  Figure 12 shows this comparison at 250 $\mu$m for fits based on 4.5-24 $\mu$m 
data  only and for fits based on 4.5-70 $\mu$m data, in both cases restricting to sources detected with at least 
5-$\sigma$ at 250 $\mu$m
and excluding sources with infrared SEDs dominated by an AGN dust torus.  The 4.5-70 $\mu$m fits show some correlation, 
with a tendency to underestimate the fluxes because of the failure to account for colder dust.  
The average value of $log_{10} (S(250)_{obs}/S(250)_{pred})$ is 0.075, corresponding to a mean underestimate
by a factor 1.2, with an rms  scatter of 0.38 dex.
Predictions based on 4.5-24 $\mu$m only data show a larger scatter compared with the observed fluxes. 
The average value of $log_{10} (S(250)_{obs}/S(250)_{pred})$ is 0.37, corresponding to a larger mean
 underestimate by a factor 2.35, with an rms  scatter of 0.51 dex.  This reflects the fact that there was no possibility
of predicting the presence of 10-20 K dust from observations at 4.5-24 $\mu$m.  Note that these mean ratios 
are biassed by the observed 250 $\mu$m flux-limit.  The correlation is much worse at 500 $\mu$m.
Elbaz et al (2010) address a slightly different issue, how well the bolometric luminosity, derived at 100-500 $\mu$m, 
is correlated with the 24 $\mu$ luminosity.  They found a good correlation, but also that 250 and
350 $\mu$m monochromatic luminosities deviated (on the high side) from their template predictions (their Fig 2).  
Our explanation for that deviation is the presence of the new cold dust components. 

\begin{figure}
\epsfig{file=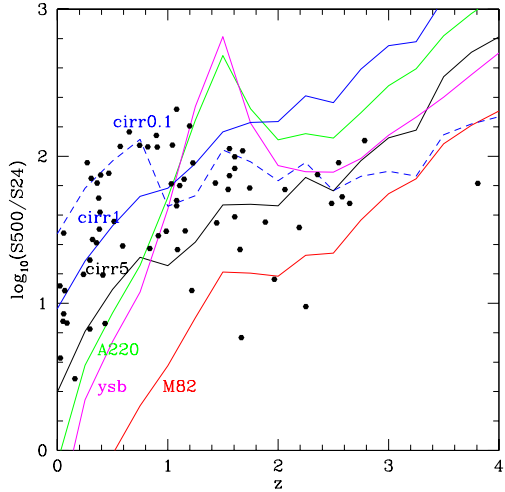,angle=0,width=7cm}
\caption{The (500/24) $\mu$m flux-ratio versus z.  Black filled circles: SPIRE-Lockman sources detected at better than 5-$\sigma$
at 250, 350 and 500 $\mu$m.  Continuous loci: predictions of infrared templates: black: cirrus ($\psi$=5); red: M82
starburst; green: A220 starburst; blue: cooler cirrus ($\psi$=1); blue broken line: colder cirrus ($\psi$=0.1); magenta: 
young starburst ($A_V$=150, t=0).
}
\end{figure}

\begin{figure*}
\epsfig{file=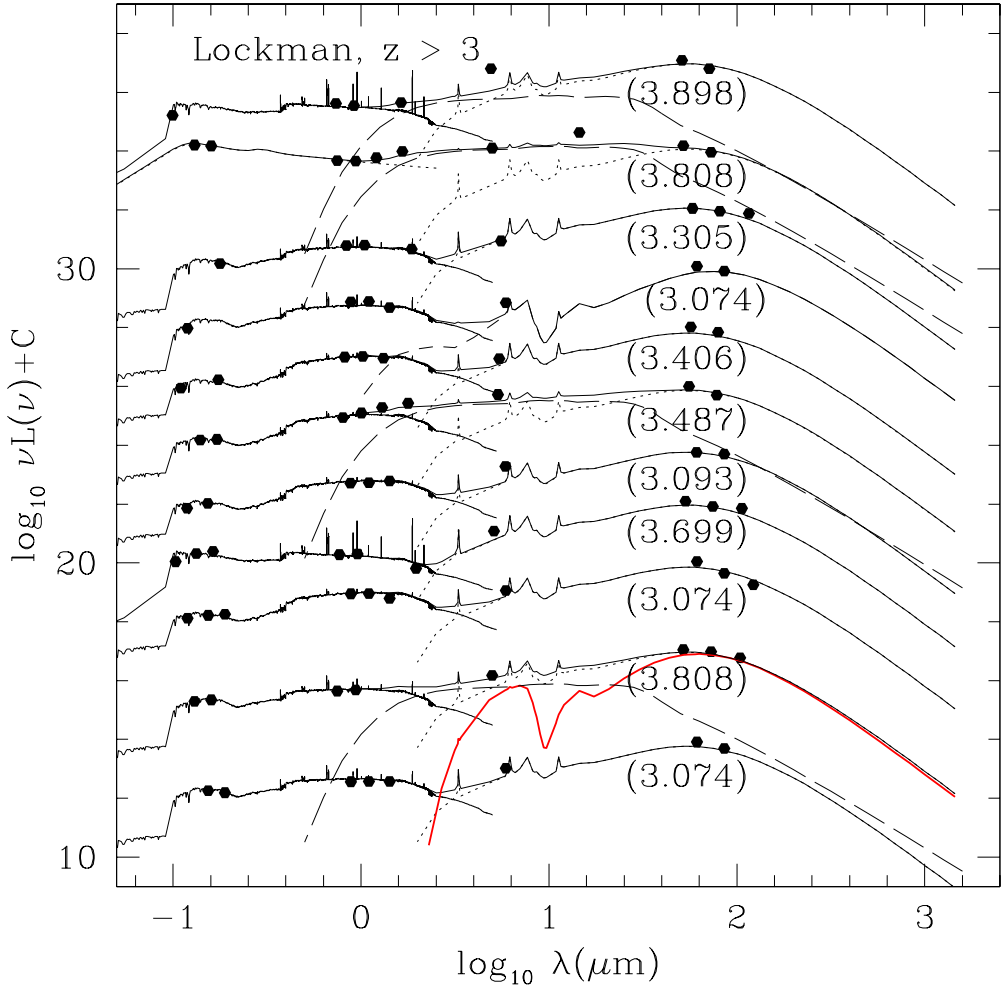,angle=0,width=14cm}
\caption{SEDs for SWIRE-Lockman 250+350 $\mu$m galaxies with {\it z}$>$ 3. 
}
\end{figure*}

\section{Discussion}
By combining {\it Herschel} 250-500 $\mu$m data with {\it Spitzer} 3.6-160 $\mu$m data we have demonstrated the
presence of two new infrared components in galaxy SEDs: a colder quiescent ('cirrus') component, and a very
young starburst component.  Local {\it IRAS} galaxies with colder cirrus than our standard $\psi$ = 5 cirrus template were discussed 
previously by Rowan-Robinson (1992).  Cooler dust was also inferred from {\it ISO} 200 $\mu$m mapping of 8 nearby
galaxies by Alton et al (1998).    They inferred a grain temperature of 18-21 K for this extended, colder component.
{\it ISO} 175/90 $\mu$m versus {\it z} (Rowan-Robinson et al 2004) and Spitzer 160/70 $\mu$m versus {\it z}
(Symeonidis et al 2009) diagrams can be interpreted as pointing to cooler dust in galaxies at {\it z}$<$ 0.4.
Here we find cooler dust to be present to a much higher range of luminosities  ($10^{12} L_{\odot}$) and redshift (1.7). 
Cold dust could be present in high redshift (z $>$ 3) galaxies but would be observable only at wavelengths $> 500 \mu$m.
The possibility that a significant fraction of galaxies detected at 850 $\mu$m could be fitted by a cirrus template was highlighted
by Efstathiou and Rowan-Robinson (2003).  The new colder cirrus templates we are using here have dust grains at 
temperatures 10-20 K for different grain types (see Table 1).  Colder dust implies lower surface brightness illumination,
and therefore more extended emission than the standard templates.

\begin{figure}
\epsfig{file=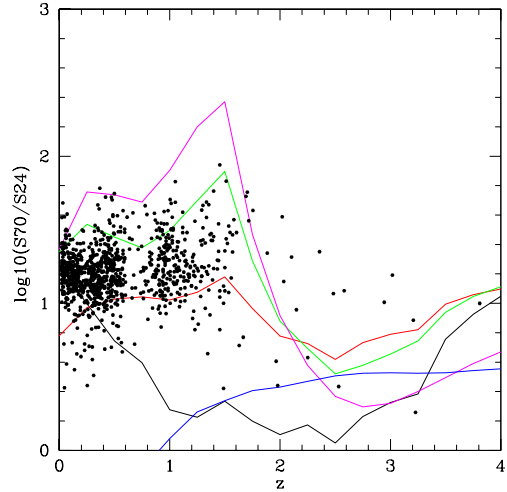,angle=0,width=7cm}
\caption{ [70/24] versus z.  Filled circles: SPIRE-Lockman sources detected at better than 5-$\sigma$ at 70, 250 and 
350 $\mu$m.  Colour-coding for model loci as in Fig 1.
}
\end{figure}

\begin{figure}
\epsfig{file=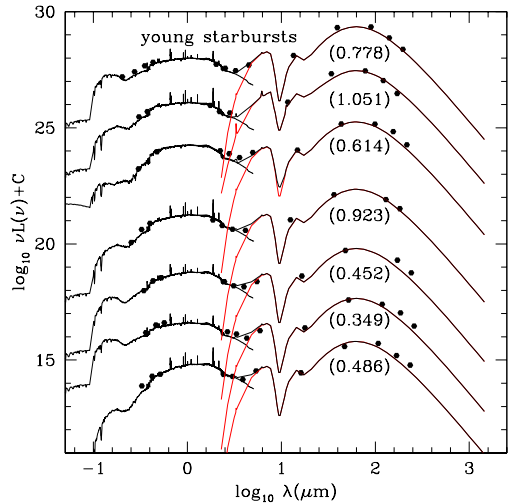,angle=0,width=7cm}
\caption{SEDs for SPIRE-Lockman galaxies with [70/24] $>$ 1.5.  Red curves are
models for very young starbursts.
}
\end{figure}

\begin{figure}
\epsfig{file=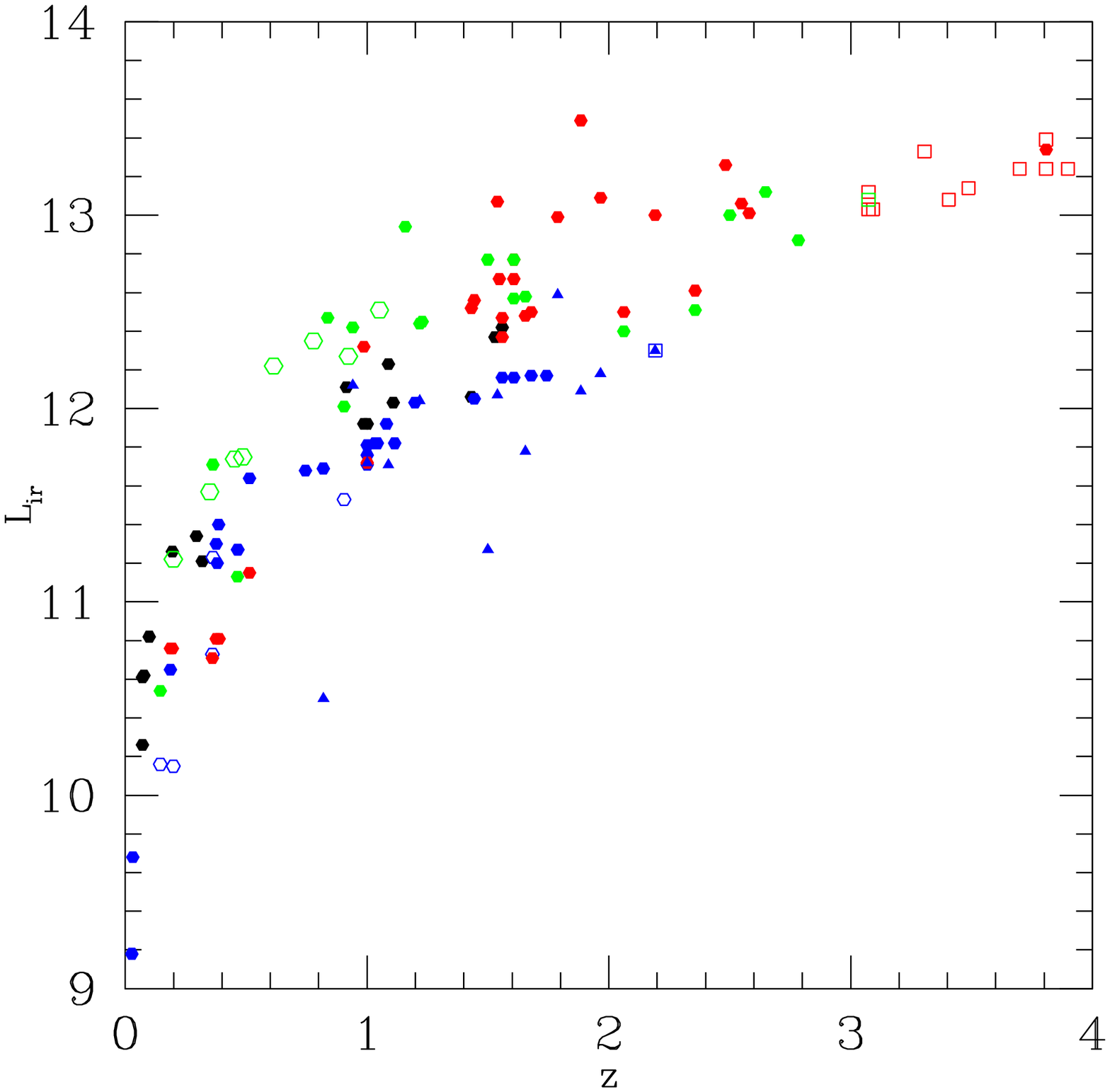,angle=0,width=7cm}
\caption{Infrared luminosities of different components as function of redshift.  Filled black hexagons: cirrus ($\psi$=5), filled blue
hexagons: cirrus ($\psi$=1), open blue hexagons: cirrus ($\psi$=0.1), filled red hexagons: M82 starbursts, filled green hexagons: Arp220 starbursts,
open green hexagons: young starbursts, open red squares: additional z$>$3 M82 starbursts, open green square: additional z$>$3Arp220
starburst, filled blue triangles: AGN dust tori.
}
\end{figure}

We have estimated the dust and stellar masses for these 68 galaxies, using the prescriptions of Rowan-Robinson et al (2008).  The star-formation
histories used to generate the optical templates yield the ratio of bolometric (or monochromatic) luminosity to stellar mass at z = 0.  
Rowan-Robinson et al (2008) give a simple prescription to correct this ratio at earlier times for the effects of passive stellar evolution. 
The  radiative transfer models for the infrared templates predict the ratio of bolometric infrared luminosity to dust mass.
Figure 13 shows dust mass versus stellar mass
for {\it Herschel} galaxies compared with the distribution for Spitzer-SWIRE galaxies.  All the galaxies with $M_{dust} < 3\times 10^8 M_{\odot}$
have {\it z}$<$0.3.  Apart from these low redshift, low luminosity, lower dust mass galaxies, Herschel 500 $\mu$m galaxies tend to be 
those galaxies with the very highest dust masses amongst the galaxies detected by Spitzer.  Assuming 
standard gas-to-dust ratios, they must have exceptionally high ratios of gas mass to stellar mass.

\begin{figure}
\epsfig{file=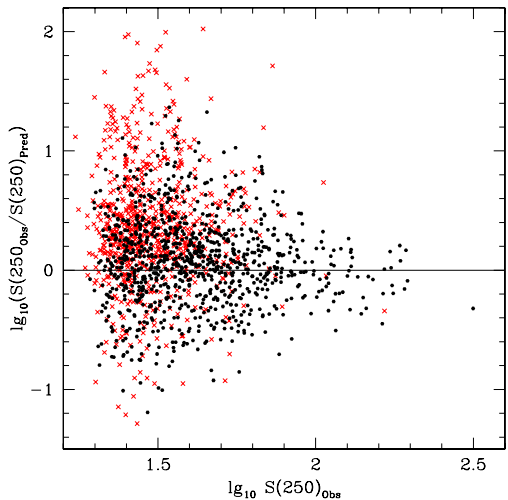,angle=0,width=7cm}
\caption{Ratio of observed flux at 250 $\mu$m flux to predicted flux, based on 4.5-24 $\mu$m data, 
 versus observed 250 $\mu$m flux (red crosses).  Filled black circles are predictions based on 4.5-70 $\mu$m data.
}
\end{figure}

\begin{figure}
\epsfig{file=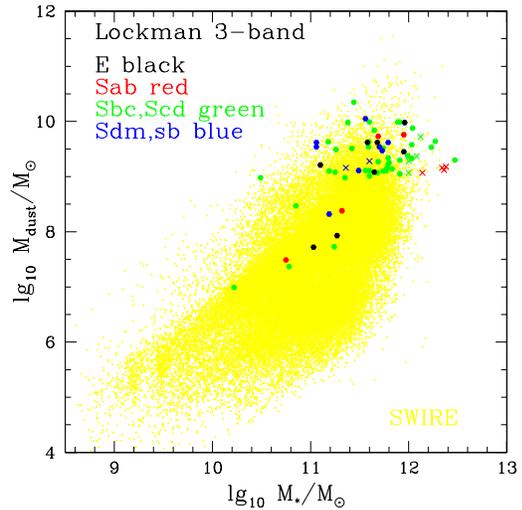,angle=0,width=7cm}
\caption{Dust mass versus stellar mass for {\it Herschel}-SPIRE galaxies (large filled circles), compared
with distribution for SWIRE Photometric Catalogue (small yellow circles).
}
\end{figure}

At low redshifts (z$<$0.2) and luminosities ($L_{ir}<10^{11} L_{\odot}$), we find $L{ir} < L_{opt}$, consistent with emission 
from an optically thin interstellar medium.  However at higher redshifts and luminosities we find cirrus components with $L_{ir}/L_{opt}$ ranging up to 
5, suggesting an optical depth in cold dust $\tau_{uv} $ at least 1-2.  For 41 of our 68 galaxies we infer dust extinctions in the range 
$A_V$ = 0.2-2, corresponding to an ultraviolet (1000 $\AA$) optical depth $\tau_{uv} \sim$ 1-10 for normal Galactic dust.  This is consistent 
with the values deduced by
Buat et al (2010) from comparison of SPIRE and Galex data.  However we also have 9 examples of galaxies at {\it z}= 0.4-1.1 with
cirrus components with high $L_{ir}/L_{opt}$, but in which
the optical starlight appears unreddened, including 5 fitted with elliptical galaxy templates in the optical.
We have to presume illumination of the cold dust is not by the stars contributing to the optical continuum, but by an obscured stellar population.  
For normal Galactic dust, the spatial scale of the dust must be tens of kiloparsecs.  A Spitzer population of elliptical galaxies with $L_{ir} > L_{opt}$ 
was discussed by Rowan-Robinson et al (2008, see Fig 20).  The optical and infrared components seem to be
unconnected with each other, perhaps reflecting a galaxy merger.  Another possibility would be that the optical galaxy is lensing a background submillimetre galaxy. 

Very young starbursts were invoked by Rowan-Robinson and Efstathiou (2009) to understand the galaxies with deepest 
silicate absorptions in their IRS spectra.  The combination of {it Herschel} and {\it Spitzer} data allow us to identify very young starbursts
from their spectral energy distribution, 
even where we do not have detailed mid-infrared spectroscopy, and hold out the prospect of determining
an age sequence among starbursts.  So far we have identified these only to z $\sim$ 1, but this is a selection effect
due to the need to have a SWIRE 70 $\mu$m detection.  Larger samples at z $>$ 2, and photometry over the full
range of wavelength from 70 to 1100 $\mu$m, will be needed to characterize the evolution of these different populations.

The range of infrared templates required to understand the 4.5-500 $\mu$m SEDs, with a range of factor 3.8 in their
peak wavelength, makes it hard to determine useful redshifts from submm data alone.  However sources whose SEDs
peak at 350 or 500 $\mu$m are likely to have {\it z}$>$ 1.

The requirement of association of our sample with the SWIRE photometric catalogue biases our sample
against higher redshift ( $>$ 1.5) since many higher redshift galaxies would be too weak at 24 $\mu$m to be
in the SWIRE sample or too faint optically to acquire photometric redshifts.  
Our sample contains 23 galaxies with {\it z}$>$ 1.5: the true number in a complete sample would be $\sim$80.
Even so we confirm the result of
Clements et al (2008) that there is a strong tail in the redshift distribution of submillimetre galaxies to lower z.

\section{Acknowledgements}
SPIRE has been developed by a consortium of institutes led by
Cardiff Univ. (UK) and including Univ. Lethbridge (Canada);
NAOC (China); CEA, LAM (France); IFSI, Univ. Padua (Italy);
IAC (Spain); Stockholm Observatory (Sweden); Imperial College
London, RAL, UCL-MSSL, UKATC, Univ. Sussex (UK); Caltech, JPL,
NHSC, Univ. Colorado (USA). This development has been supported
by national funding agencies: CSA (Canada); NAOC (China); CEA,
CNES, CNRS (France); ASI (Italy); MCINN (Spain); SNSB (Sweden);
STFC (UK); and NASA (USA).

The data presented in this paper will be released through the {\it Herschel}  
Database in Marseille HeDaM\footnote{hedam.oamp.fr/HerMES}.

%\clearpage

\begin{table*}
\caption{Parameters for SED models, complete sample of 68 3-band sources
(luminosities and dust masses are in $log_{10}$ solar units)}
\begin{tabular}{llllllllllll}
RA & dec & {\it z}&  $L_{cirr, \psi =5}$ &  $L_{cirr, \psi =1}$ & $L_{sb, M82}$ &  $L_{sb, A220}$   &$L_{tor}$ & $L_{opt}$ & template & $A_V$& $M_{dust}$\\
 162.94820 & 58.26883 & 0.0280 & & 9.18 & & &  & 9.79 & Scd & 0.25 & 6.99\\
 161.11473 & 58.90322 & 0.0310 & & 9.68 & & &  & 10.19 & Sab & 0.0 & 7.49\\
 159.98526 & 57.40519 & 0.0720 & 10.26 & & & &  & 10.36 & Scd & 0.2 & 7.37\\
 160.10109 &  58.15504 &  0.0721 & 10.61 & & & &  & 10.46 & E & 0.0 & 7.72\\
 164.36295 & 57.94135 & 0.0770 & 10.62 & & & &  & 10.82 & Scd & 0.3 & 7.73\\
 160.46895 & 58.97295 & 0.1000 & 10.82 & & & &  & 10.92 & Sab & 0.4 & 7.93\\
 162.21977 & 59.63440 & 0.1449 & & 10.16* & & 10.54 &  & 10.09 & Scd & 0.0 & 8.98\\
 161.11208 &  57.70857 &  0.1878 & & 10.65 & 10.76 &  & & 10.46 & Scd & 0.2 & 8.47\\
 158.99170 &  58.97944 &  0.1957 & 11.26 & & 10.76 &  & & 10.81 & Sab & 0.5 & 8.38\\
 161.34665 &  57.51625 &  0.1991 & & 10.15* & &  11.22** & & 10.96 & Scd & 0.3 & 8.98\\
 162.41336 & 57.68822 & (0.294) & 11.34 & & & &  & 11.23 & sb & 0.0 & 8.32\\
 159.88918  & 59.07882 & (0.318) & 11.21 & & &  & & 11.23 & Scd & 0.0 & 9.54\\
 160.00403 &  57.31637 &  (0.361) & & 10.73* & 10.71 &  & & 11.11 & sb & 0.8 & 9.54\\
 162.80595 &  57.24055 &  0.3620 & & 11.23* & & 11.71 &  & 11.61 & sb & 1.9 & 10.05\\
 160.35732 &  59.34703 &  0.3770 & & 11.30 & 10.81 & &  & 11.26 & Scd & 1.0 & 9.11\\
 162.35535 &  57.92122 &  (0.380) & & 11.20 & 10.81 &  & & 11.26 & Scd &1.0 & 9.01\\
 161.11140 &  59.45845 & (0.387) & & 11.40 & & &  & 10.62 & E & 0.0 & 9.21\\
 162.11961 &  58.27710 & 0.4635 & & 11.27 & & 11.13 &  & 10.87 &Scd & 0.75 & 9.10\\
 163.17931 & 58.67978 & (0.466) & & 11.27 & & &  & 10.93 & Scd & 1.8 & 9.08\\
 161.72240 & 59.65389 & (0.514) & & 11.64 & 11.15 &  & & 11.50 & E & 0.0 & 9.45\\
 161.32074 &  58.78179 &  (0.746) & & 11.68 & & &  & 10.99 & Scd & 0.8 & 9.49\\
 162.54356 &  57.91158 &  (0.820) & & 11.69 & &  & 10.50 & 11.60 & QSO & 0.2 & 9.50\\
 160.39917 &  58.65377 &  (0.837) & & & & 12.47 &  & 11.32 & Sbc & 0.0 & 9.11\\
 160.97595 &  59.36114 &  (0.905) & &11.53* & & 12.01  & & 11.21 & Sbc & 0.6 & 10.35\\
161.62852 &  57.54497 &  (0.914) & 12.11 & & & & & 11.56 & Sbc & 0.2 & 9.22\\
 160.51041 & 58.67371 & 0.9416 & & & & 12.42 &  12.12 & 12.12 & QSO & 0.8 & 9.06\\
 161.28079 & 57.89500 & (0.986) &11.92 & & 12.32 &  & & 11.55 & Sbc & 0.2 & 9.11\\
 160.76065 &  58.44792 &  (1.0) & & 11.81 & & &  & 11.22 & E & 0.0 & 9.62\\
 162.94107 & 58.85802 & (1.0) & & 11.81 & & &  & 11.32 & E & 0.0 & 9.62\\
 161.44548 & 59.58356 & (1.0) & & 11.71 & & &  & 11.52 & Scd & 0.5 & 9.52\\
 159.70134 & 58.61901 & (1.0) & 11.92 &  & 11.72 &  & 11.72 & 11.92 & QSO & 0.3 & 9.05\\
 163.31041 & 58.25547 & (1.032) & & 11.81 & & &  & 11.22 & sb & 0.15 & 9.62\\
 158.92084 & 57.47687 & (1.042) & & 11.82 & & & & 11.38 & Sbc & 0.4 & 9.63\\
 159.27916 &  57.51514 &  (1.080) & & 12.17 & &  & & 11.62 & E & 0.0 & 9.98\\
 162.93555 & 58.62388 & (1.080) & & 11.92 & &  & & 11.38 & Sab & 0.0 & 9.73\\
 159.02222 &  59.17973 & (1.089) & 12.23 & & &  & 11.71 & 11.59 & Scd & 0.3 & 9.34\\
 159.12962 &  58.90953 & (1.109) & 12.03 & & & & & 11.63 & Scd & 0.3 & 9.14\\
 159.05000 & 58.45025 & (1.115) & & 11.82 & & & & 10.98 & Sbc & 0.0 & 9.63\\
 162.70114 &  57.71767 & (1.158) & & & & 12.94 &  & 11.84 & Scd & 0.2 & 9.58\\
 159.70450 & 58.38959 & (1.198) & & 12.03 & & &  & 11.46 & Sbc & 0.4 & 9.84\\
 162.19649 & 57.39379 & (1.218) & & & & 12.44  & 12.04 & 11.33 & E & 0.0 & 9.08\\
 161.19606 & 57.60232 & (1.228) & & & & 12.45  & & 11.50 & Scd & 0.0 & 9.07\\
 159.49042 &  57.76735 &  (1.432) & 12.06 & & 12.52 &  & & 11.65 & Scd & 0.45 & 9.25\\
 161.24333 & 57.55045 & (1.443) & & 12.05 & 12.56 &  & & 11.89 & Scd & 0.25 & 9.88\\
 162.01488 & 58.90593 & (1.5) & & &  & 12.77 &  11.27 & 11.17 & QSO & 0.6 & 9.39\\
 158.80722 &  57.57902 &  1.5386 & & & 13.07  & & 12.07 & 11.72 & sb & 0.2 & 9.11\\
 163.39421 & 57.71170 & (1.547) & 12.37 & & 12.67 &  & & 11.93 & sb & 0.8 & 9.54\\
 160.05550  & 58.52735 & (1.559) & & 12.16 & 12.37  & & & 11.25 & Scd & 0.2 & 9.98\\
 163.81096 & 57.99709 & (1.559) & 12.47 & & 12.47 & &  & 12.02 & sb & 1.0 & 9.62\\
 161.86531 &  57.94479 &  (1.606) & & & 12.57 & 12.77  & & 11.97 & sb & 0.9 & 9.47\\
 164.09378 & 58.25208 & (1.606) & & 12.16 & 12.67 &  & & 11.78 & Sbc & 0.2 & 9.99\\
 159.06248 & 57.91819 & (1.606) &  & & 12.67  & 12.57 & & 11.90 & Sbc & 0.1 & 9.33\\
 163.18320 &  58.37976 &  (1.655) & & & 12.48  & 12.58 & 11.78 & 11.78 & Sbc & 0.1 & 9.30\\
 160.54668 & 59.17580 & (1.679) & & 12.17 & 12.48 & &  & 11.48 & Scd & 0.4 & 9.99\\
 162.10071 & 59.34946 & (1.742) & & 12.17 & 12.48 & &  & 11.78 & Scd & 0.8 & 9.99\\
 162.33145 & 58.98063 & 1.7880 & & & 12.99 & &  12.59 & 12.69 & QSO & 0.1 & 9.03\\
 162.91730 &  58.80596 &  (1.884) & & & 13.49 &  &  12.09 & 12.13 & Sbc & 0.1 & 9.53\\
 161.54591 &  58.16109 & (1.965) & & & 13.09 &  & 12.18 & 12.56 & QSO & 0.0 & 9.13\\
&&&&* $\psi$=0.1 &&** young sb&&&&&\\
\end{tabular}
\end{table*}

\begin{table*}
\caption{Parameters for SED models, contd.}
\begin{tabular}{llllllllllll}
RA & dec & {\it z}&  $L_{cirr, \psi =5}$ &  $L_{cirr, \psi =1}$ & $L_{sb, M82}$ &  $L_{sb, A220}$   &$L_{tor}$ & $L_{opt}$ & template & $A_V$& $M_{dust}$\\

 159.22012 & 58.31201 & (2.062) & & & 12.50 & 12.40 &  & 11.70 & Sbc & 0.1 & 9.16\\
 162.36497 &  59.06819 & (2.192) & & & 13.00 & &  12.30 & 12.70 & QSO & 0.0 & 9.04\\
 164.85812 &  58.30039 & (2.357) & & & 12.61 & 12.51   & & 11.65 & Scd & 0.4 & 9.27\\
 159.61342  & 57.87177 & (2.483) & & & 13.26 & & &  12.45 & Scd & 0.55 & 9.30\\
 163.17749 &  58.66597 &  (2.5) & & & & 13.00 & &  12.25 & Scd & 0.5 & 9.64\\
 162.84056 &  57.70921 &  2.5480 & & & 13.06 &  & & 11.71 & Sbc & 0.0 & 9.10\\
 161.54903 &  59.04079 &  (2.581) & & & 13.01  & & & 11.90 & Scd & 0.5 & 9.05\\
 161.45622 & 57.56762 & (2.648) & & & & 13.12 &  & 11.85 & Sab & 0.0 & 9.76\\
 161.85199 & 59.06105 & (2.784) & & & & 12.87 &  & 11.41 & Sbc & 0.0 & 9.51\\
 163.07271 & 58.98540 & (3.808) & & & 13.34 & &  & 12.08 & Scd & 0.0 & 9.38\\
&&&&&&&&&&&\\
young sb &&&&&&&&&&&\\
  161.65623  & 57.73351 & (0.486) & & & &  11.75** & & 11.04 & Sbc & 1.0 & 8.27\\
  162.53716  & 57.82069 & (0.349) & & & &  11.57** & & 10.81 & Scd & 0.4 & 8.09\\
  159.67081 & 57.94089 & (0.452) & & & &  11.74** & & 11.07 & Scd & 0.8 & 8.26\\
  161.55408 & 58.09828 & (0.923) & & & &  12.37** & & 11.45 & Sbc & 0.3 & 8.89\\
  159.75601 & 58.53660 & (0.614) & & & &  12.22** & & 11.32 & E & 0.0 & 8.74\\
  163.56523 & 58.70536 & (1.051) & & & &  12.51*** & & 11.26 & Sbc & 0.0 & 9.03\\
  160.28583 & 58.77047 & (0.778) & & & &  12.35** & & 11.29 & Scd & 0.4 & 8.87\\
&&&&&&** young sb, t=0&&&&&\\
&&&&&&*** ysb, t=6 Gyr&&&&&\\
&&&&&&&&&&&\\
z $>$ 3 &&&&&&&&&&&\\
164.67636  & 57.55608 & (3.074) & & & 13.03 & & & 12.03 & Scd & 0.0 & 9.07\\
159.28676  & 57.65699  & (3.808) & & & 13.24 & & 12.24 & 12.09 & Scd & 0.0 & 9.28\\
& & & & & & 13.05** &  12.24 & 12.09 & Scd & 0.0 & 9.57\\
159.18820  & 58.03481 & (3.074) & & & 13.12 & & & 12.28 & Sab & 0.0 & 9.16\\
164.53999  & 58.03517 & (3.699) & & & 13.24 & & & 11.79 & sb & 0.0 & 9.28\\
161.60800  & 58.23610 & (3.093) & & & 13.03 & & & 12.08 & Sab & 0.1 & 9.07\\
163.42009  & 58.61931 & (3.487) & & & 13.14 & & 12.88 & 12.36 & Sab & 0.1 & 9.18\\
162.51204  & 58.11813 & (3.406) & & & 13.08 & &  & 12.33 & Sab & 0.0 & 9.12\\
163.03342 &  58.16816 & (3.074) & & & & 13.08 & & 12.15 & Scd & 0.15 & 9.72 \\
164.52054  & 58.30782 & (3.305) & & & 13.33 & & & 12.13 & Scd & 0.1 & 9.37\\
159.67590  & 59.01916 & (3.808) & & & 13.34 & & 13.52 & 13.52 & QSO & 0.0 & 9.38\\
161.78809  & 58.72614 & (3.898) & & & 13.24 & &  12.24 & 12.04 & sb & 0.0 & 9.28\\
\end{tabular}
\end{table*}

\end{document}

%% file: macro.tex
\newcommand{\aip}{{\small ${\cal AIPS}$}}
\newcommand{\gtsim}{\mbox{{\raisebox{-0.4ex}{$\stackrel{>}{{\scriptstyle\sim}}
$}}}}
\newcommand{\ltsim}{\mbox{{\raisebox{-0.4ex}{$\stackrel{<}{{\scriptstyle\sim}}
$}}}}
\newcommand{\s}{$\stackrel{\rm s}{.}$}
\newcommand{\h}{$^{\rm h}$}
\newcommand{\m}{$^{\rm m}$}
\newcommand{\pp}{$\stackrel{\prime\prime}{.}$}
\newcommand{\de}{$^{\circ}$}
\newcommand{\p}{$^{\prime}$}
\newcommand{\arc}{$^{\prime\prime}$}
\newcommand{\marc}{^{\prime\prime}}
\newcommand{\rs}{{\em $r_s$}}
\newcommand{\DPM}{{\em DPM}}
\newcommand{\alf}{{\displaystyle\biggl({\nu_{\rm h} \over \nu_{\rm l}}\biggr)^{\alpha}} }

\newcommand{\figstart}[1]
    { \begin{figure}[htb]
      \begin{picture}(0,#1) }
\newcommand{\figend}[4]
    { \end{picture}
      \special{#1}
      \caption[#2]{#3}
      \label{#4}
      \end{figure} }
\newcommand{\fig}[5]
    { \figstart{#1}
      \figend{#2}{#3}{#4}{#5} }
\newcommand{\bHS}{\beta_{\mbox{\scriptsize HS}}}
\newcommand{\bBF}{\beta_{\mbox{\scriptsize BF}}}
\newcommand{\nT}{\nu_{\mbox{\scriptsize T}}}
\newcommand{\et}{E_{\mbox{\scriptsize T}}}
\newcommand{\nTn}{\nu_{\mbox{\scriptsize Tn}}}
\newcommand{\nTf}{\nu_{\mbox{\scriptsize Tf}}}
\newcommand{\tn}{\tau_{x\mbox{\scriptsize n}}}
\newcommand{\tf}{\tau_{x\mbox{\scriptsize f}}}
\newcommand{\xn}{x_{\mbox{\scriptsize n}}}
\newcommand{\xf}{x_{\mbox{\scriptsize f}}}
\newcommand{\yn}{y_{\mbox{\scriptsize n}}}
\newcommand{\yf}{y_{\mbox{\scriptsize f}}}
\newcommand{\lln}{l_{\mbox{\scriptsize n}}}
\newcommand{\llf}{l_{\mbox{\scriptsize f}}}
\newcommand{\Dn}{f(\Delta_{\mbox{\scriptsize n}})}
\newcommand{\Df}{f(\Delta_{\mbox{\scriptsize f}})}
\newcommand{\B}{\mbox{$B$}}
\newcommand{\Bo}{\mbox{$B$}_{0}}